\documentclass[twocolumn,showpacs,superscriptaddress,preprintnumbers,amsmath,amssymb]{revtex4}
\usepackage{amsmath}
\usepackage{graphicx}

\begin{document}

\title{Nonlinear resonant behavior of the dispersive readout scheme for a superconducting flux qubit}

\author{Janice C. Lee}
 \affiliation{Department of Electrical Engineering and Computer Science,\\
  Massachusetts Institute of Technology, Cambridge, Massachusetts 02139, USA}
\author{William D. Oliver}
\affiliation{MIT Lincoln Laboratory, 44 Wood Street, Lexington,
Massachusetts 02420, USA}
 \author{Karl K. Berggren}
 \altaffiliation[Present address: ]{EECS Department, MIT} 
 \affiliation{MIT Lincoln Laboratory, 44 Wood Street, Lexington, Massachusetts 02420, USA}
\author{T. P. Orlando}
 \affiliation{Department of Electrical Engineering and Computer Science,\\
 Massachusetts Institute of Technology, Cambridge, Massachusetts 02139, USA}
\date{\today}

\begin{abstract}

A nonlinear resonant circuit comprising a SQUID magnetometer and a
parallel capacitor is studied as a readout scheme for a
persistent-current (PC) qubit. The flux state of the qubit is
detected as a change in the Josephson inductance of the SQUID
magnetometer, which in turn mediates a shift in the resonance
frequency of the readout circuit.  The nonlinearity and resulting
hysteresis in the resonant behavior are characterized as a
function of the power of both the input drive and the associated
resonance peak response. Numerical simulations based on a
phenomenological circuit model are presented which display the
features of the observed nonlinearity.
\\
\end{abstract}

\maketitle\pagebreak

\section{Introduction}

Superconducting Josephson junction circuits are promising
candidates for realizing a quantum computer. These solid-state
qubits can be fabricated using standard integrated-circuit
techniques, where there is the possibility to incorporate the
control and readout circuitry on-chip, providing a manageable
option for scaling up to a larger number of qubits.
Quantum-coherent phenomena~\cite{Makhlin01a} have been studied
utilizing the quantum states~\cite{Clarke88} of single-qubit
circuits and cavities, including superpositions of distinct
macroscopic states~\cite{Friedman00a,Wal00a}, time-dependent Rabi
oscillations~\cite{Naka1,Irinel,Saito,Rabi_Clarke,Yang1,
Martinis_Rabi,Dion}, cavity
quantum-electrodynamics~\cite{Chiorescu04,Wallraff04,Johansson06}
and Mach-Zehnder-type interferometry
\cite{Oliver05,Sillanpaa06,Berns06}. Coherent
oscillations~\cite{Naka2}, spectroscopic evidence for
entanglement~\cite{Berkley}, and a prototypical gate
operation~\cite{Naka3} have also been demonstrated in
superconducting coupled-qubits. However, to further increase the
coherence times of these qubits for manipulation of their quantum
states, one must find ways to reduce the amount of noise intrinsic
to the qubit, as well as noise introduced by the readout process
itself. In particular, several previous readout methods have
relied on the switching of a Josephson circuit from a zero-voltage
to a finite-voltage state. This switching generates
quasiparticles, and thus such readout approach is limited by the
subsequent decoherence. More recently dispersive readout schemes
have been developed such that the qubit is coupled to a resonator,
and the state of the qubit is detected as a shift in the resonance
frequency of the resonator. As a result, the readout process
requires only lower input biases and hence minimizes the
generation of quasiparticles. Furthermore, the resonator also acts
as a narrow-band filter which shields the qubit from broadband
noise. Dispersive readout has been implemented for the persistent
current qubit \cite{Adrian,Ilichev1}, for the charge qubit
\cite{Wallraff}, and for the hybrid qubit where the readout was
operated in the nonlinear regime for its use as a bifurcation
amplifier \cite{Siddiqi}.

\begin{figure}[ht]
   \begin{center}
   \includegraphics[width=3.5in]{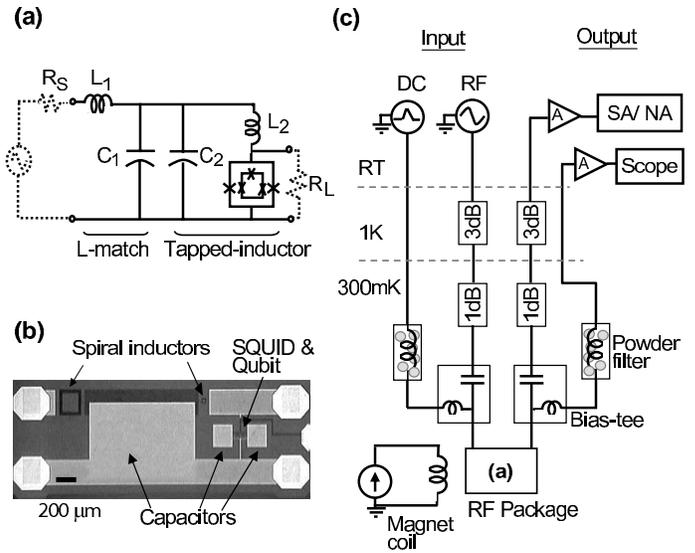}\\
   \caption{Experimental setup: (a) Circuit schematic of the resonant
   readout circuit.  The designed component values were $L_1$ = 69\,nH,
   $L_2$ = 0.78\,nH, $C_1$ = 1.4\,pF, and $C_2$ = 100\,pF.
   The SQUID inductance $L_J$ was approximated to be 0.2\,nH for the
   circuit design. (b) Optical micrograph of the actual device. (c)
   Electronic setup at different temperature stages of the $^3{\rm He}$ cryostat.}
   \label{fig:Fig1_Scheme}
   \end{center}
\end{figure}

This paper focuses on characterizing the nonlinear resonant
behavior of the dispersive readout scheme developed for a
persistent current qubit.  The readout element is a SQUID
magnetometer, which is operated as a nonlinear, flux-sensitive
inductor incorporated in an L-C resonator.  The qubit is coupled
to the SQUID inductor, and the flux state of the qubit is detected
as a shift in the resonance frequency of the resonator by means of
magnitude and/or phase measurements. Our approach differs from
other resonant-type experiments in two main ways. First, our qubit
and readout circuit were fabricated on the same chip from niobium,
whereas the implementations in \cite{Adrian}-\cite{Siddiqi} were
aluminum-based. Second, we were able to achieve a high quality
factor for the resonator by incorporating an RF transformation
network on-chip using the planarized niobium process. We observed
resonant behavior due to the nonlinear Josephson inductance of the
SQUID, given the high quality factor of the resonance
\cite{abdo,Oates}. The frequency spectra of the readout circuit
were characterized in both the linear and nonlinear regimes.
Biasing the readout circuit in the nonlinear regime potentially
provides additional sensitivity for distinguishing the qubit
states \cite{Siddiqi}.

The persistent current (PC) qubit used in this study is a
superconducting loop interrupted by three Josephson junctions, two
of which have the same critical current while the third junction
has a critical current reduced by a factor $\alpha$
\cite{Mooij_Sci,TPOPRB}.  When the external magnetic flux
threading the qubit loop is biased near half a flux quantum, the
two lowest energy states correspond to oppositely circulating
persistent currents in the qubit loop. The induced flux of the
persistent current (and hence the state of the qubit) is detected
by a SQUID magnetometer which surrounds the qubit.

In the resonant readout scheme, the SQUID magnetometer is operated
using the property that the Josephson inductance of the SQUID is a
nonlinear function of both the current bias $I_{sq}$ and the flux
bias $\Phi_{ext}$.  In our experiments, the SQUID current bias
comprises solely an AC component, whereas the flux bias
$\Phi_{ext} = \Phi_{dc} + \Phi_{ac}$ has both a DC component
corresponding to the external bias, and an AC component
corresponding to the induced flux that is mutually coupled to the
SQUID. To demonstrate the general principles underlying the
operation, consider the limiting case where the SQUID has
negligible loop inductance and symmetric junctions each with
critical current $I_{co}$.  In this limit, the SQUID behaves like
an equivalent single junction with an effective critical current
given by $I_c(\Phi_{ext})=2I_{co}|\cos(\pi\Phi_{ext}/\Phi_o)|$,
and an effective phase given by
$\varphi_p=\sin^{-1}(I_{sq}/{I_c(\Phi_{ext})})$. By defining the
Josephson inductance according to $V_{sq}=d\left[L_J
I_{sq}\right]/dt$, where $V_{sq}$ is the voltage across the SQUID,
we obtain the inductance to be
\begin{equation}
\label{eq:LJ_SQUID} L_J(I_{sq},\Phi_{ext}) =
\frac{\Phi_o}{2\pi{I_c(\Phi_{ext})}}\frac{\varphi_p}{\sin\varphi_p}\;.
\end{equation}
In the linear regime where the current and flux biases are small,
one can approximate the SQUID by a linear inductor given by
\begin{equation}
\label{eq:LJo} L_{Jo} = \frac{\Phi_o}{4\pi I_{co}}\;.
\end{equation}
The inductance for small AC drives can be approximated
quasi-statically by the inductance at the DC operating points for
the current and flux biases.  To demonstrate the separate effects
due to the current and the flux, we first set $\Phi_{ext}$ to zero
and reduce Eqn.~\ref{eq:LJ_SQUID} to
\begin{equation}
L_J(I_{sq},0) = 2L_{Jo} \frac{\varphi_p}{\sin\varphi_p}\;,
\end{equation}
where $\varphi_p=\sin^{-1}(I_{sq}/2I_{co})$.  Thus, the inductance
increases with the size of the driving DC current bias. Likewise,
when $I_{sq}$ in Eqn.~\ref{eq:LJ_SQUID} is set to zero,
\begin{equation}
L_J(0,\Phi_{ext})=\frac{L_{Jo}}{\left\vert\cos\left(\frac{\pi\Phi_{ext}}{\Phi_o}\right)\right\vert}
\end{equation}
which has a periodic dependence on the DC flux bias with
periodicity given by $\Phi_o$. Starting from a bias of
$\Phi_{dc}=0$, the inductance increases with flux, and starting
from a bias of $\Phi_{dc}=0.5\,\Phi_o$, the inductance decreases
with flux.

The general trend of an AC bias can be conceptualized as averaging
the inductance about the DC bias point over the range of the AC
bias. Hence, near $\Phi_{dc}=0$ both the AC driving current and the
AC flux increase the effective inductance as  the AC drives
increase. In contrast, near $\Phi_{dc}=0.5\,\Phi_o$ an increasing AC
flux bias tends to decrease the effective inductance and an
increasing AC current bias tends to have the opposite effect.
Therefore,the current and flux act in concert at $\Phi_{dc} = 0$;
whereas, they compete at $\Phi_{dc} = 0.5\,\Phi_o$. From our
experiments and  numerical simulations \cite{Jan_PhDthesis}, we have
found in our readout circuit that the AC flux dominates at
$\Phi_{dc} = 0.5\, \Phi_o$. Consequently, this paper will focus on
the effects due to flux, thereby allowing us to develop a
phenomenological model that qualitatively matches the experimental
observations.

Fig.~\ref{fig:Fig1_Scheme}a shows the circuit schematic of the
resonant readout circuit. The PC qubit is mutually coupled to the
SQUID inductor $L_J$.  The resonating loop comprises $L_J$, $L_2$,
and the parallel combination of $C_1$ and $C_2$. To raise the
quality factor of the resonance for higher readout sensitivity, a
tapped-inductor transformer formed by $L_2$ and $L_J$ is used to
step up the effective output resistance at the resonance
frequency. On the input side, $L_1$ and $C_1$ form an L-match
network which matches the input resistance to the transformed
output resistance \cite{Bowick}. $R_s$ and $R_L$ represent the
$50\,\Omega$ source and load impedances from the RF electronics,
and no resistors were fabricated on-chip.  The junctions of the
SQUID are each shunted by a 5 pF capacitor (not shown).  The
device was fabricated using the planarized niobium trilayer
process at MIT Lincoln Laboratory \cite{Karl}.  A device
micrograph is shown in Fig.~\ref{fig:Fig1_Scheme}b. The Josephson
critical current density was estimated to be $1.2\,\mu{\rm
A}/\mu{\rm m}^2$ from the process test data. The designed junction
dimensions were $1.0\,\mu{\rm m}$ and $0.9\,\mu{\rm m}$ for the
qubit, and $1.5\,\mu{\rm m}$ for the SQUID. Due to process bias,
the effective electrical junction dimensions are expected to be
smaller. We measured the effective size of the SQUID junctions to
be approximately $1.3\,\mu{\rm m}$, a reduction of $0.2\,\mu{\rm
m}$ from their drawn dimension. The effective qubit junction sizes
were not measured directly, but were estimated to have a reduction
of approximately $0.35\,\mu{\rm m}$, as determined by measuring
similarly drawn $1.0\,\mu{\rm m}$ process-test junctions nearby.
The area ratio of the SQUID to the qubit loop was designed to be
1.3, with mutual coupling estimated to be 30\,pH. The inductors
were realized by square spirals with a linewidth and spacing of
$1\,\mu{\rm m}$, while the capacitors comprised Nb electrodes with
a dielectric consisting of 50\,nm of Nb$_2$O$_5$ and 200\,nm of
SiO$_2$.

Our measurements were taken in a $^3{\rm He}$ cryostat at 300\,mK.
The measurement setup is shown in Fig.~\ref{fig:Fig1_Scheme}c. The
DC lines were used to characterize the junction properties, while
the RF lines were used for the resonant readout.  The external DC
flux bias for the qubit was provided by a superconducting coil
wrapped around the sample housing. The signal from the resonant
circuit was amplified at room temperature. We measured the
transmission characteristics of the readout circuit with a
spectrum analyzer equipped with a tracking generator, or with a
network analyzer when the phase information was needed.  We used a
resolution bandwidth (RBW) of 3\,kHz, and averaged each spectrum
100 times.

\section{Qubit Readout and Effect of Input Bias on Readout Circuit}

\begin{figure}[ht]
   \begin{center}
   \includegraphics[width=3.4in]{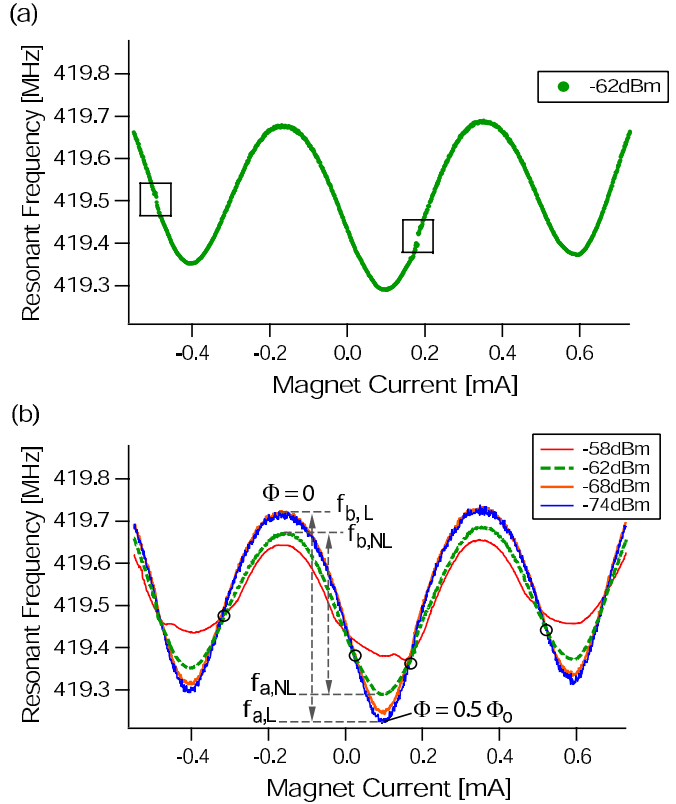}\\
   \caption{(Color online)(a) Modulation of the
  resonance frequency with external DC flux bias.  Qubit steps are observed at 0.18\,mA and -0.495\,mA.
  (b) Modulation of the resonance frequency
  for various input power.  Amount of modulation is reduced in the nonlinear regime ($f_{b,NL}-f_{a,NL}$)
  compared to the linear regime ($f_{b,L}-f_{a,L}$).
  The circular markers represent the inflection points where ${d^2f_o}/{d\Phi^2} =0$.}
   \label{fig:Fig2_modulation}
   \end{center}
\end{figure}

The resonance frequency of the readout circuit was measured to be
near 419\,MHz, with a quality factor estimated to be on the order
of 1000. Fig.~\ref{fig:Fig2_modulation}a shows the results when an
external flux bias $\Phi_{dc}$ was applied through the sample. At
a given $\Phi_{dc}$,  we measured both the resonance frequency and
the peak power of the resonance spectrum. The resonance frequency
of the readout circuit, in the linear regime where all the AC
biases are small, is related to the effective inductance $L_J$ and
capacitance $C$ by
\begin{equation}
f_o(\Phi_{dc})=\frac{1}{2\pi\sqrt{L_J(\Phi_{dc})C}} \; .
\end{equation}
A periodic modulation of the resonance frequency of the readout
circuit was observed and is interpreted as being caused by the
periodic modulation of the Josephson inductance of the SQUID. At
every 1.3 times the SQUID modulation period, a shift in the
resonance frequency, corresponding to about 2\,pH (1\%) change in
Josephson inductance, was observed. These shifts, referred to as
qubit steps, occur as the flux in the qubit $\Phi_{q}$ is swept
past $\Phi_q = 0.5\,\Phi_o$, as it is more energetically favorable
for the qubit to change from one circulating current state to the
other in order to remain in the ground state. The periodicity of
the qubit steps (corresponding to a flux quantum for the qubit)
and the periodicity of the SQUID lobes (corresponding to a flux
quantum for the SQUID) are related by the ratio of their loop
areas, which was defined by the fabrication parameters. In
addition, we observed a dip in the resonance-peak power (not
shown), which corresponds to a broadening of the resonance near
the qubit step region \cite{Jan_IEEE}. The parabolic-like
background observed in the frequency modulation curve was due to
undesired heating from the magnet current in the DC (soft-coax)
lines. The heating causes an increase in the resonance frequency,
and is more significant at high magnet current biases. The heating
effect was eliminated  for faster scans and when the sample was
later tested in a dilution refrigerator using superconducting
magnet leads.

Fig.~\ref{fig:Fig2_modulation}b shows that as the level of input
bias increases, the amount by which the frequency is modulated
over a flux quantum decreases. This will be shown in the next
section to be a direct consequence of the shape of the resonance
spectrum as it becomes increasingly nonlinear with higher input
power.

\section{Nonlinear resonant behavior of readout circuit}
\label{sect:NonlinearResBehavior}

The resonant readout circuit can experimentally  distinguish the
difference in the flux produced by the circulating current states
of the qubit.  Given that the efficacy of the readout scheme
depends on the nonlinear response of the readout circuit, we now
characterize the resonant behavior of the readout circuit as a
function of the DC flux bias for higher AC drives.

\begin{figure}[h]
   \begin{center}
   \includegraphics[width=3.5in]{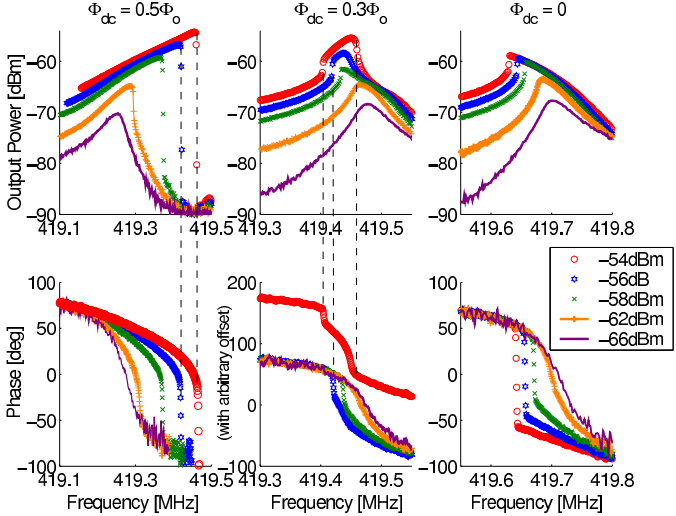}\\
   \caption
   {Evolution of the magnitude and phase spectra of the readout circuit from the linear to
   the nonlinear regime with increasing input
   power.  Data are shown for flux biases at $\Phi_{dc}=0$, $0.3\,\Phi_o$ and $0.5\,\Phi_o$.
   The nonlinear spectrum evolves from having a lower resonance
   frequency at $\Phi = 0$ to having a higher resonance frequency at
   $\Phi = 0.5\,\Phi_o$.  A self-resonance (due to parasitic couplings) was observed near the resonance frequency of the spectrum.  The phase spectrum at -54\,dBm for $\Phi
   = 0.3\,\Phi_o$ was arbitrarily shifted for display purpose.}
   \label{fig:Fig3_Mag_Phase_compareToy}
 \end{center}
\end{figure}

Fig.~\ref{fig:Fig3_Mag_Phase_compareToy} shows the evolution of
the magnitude and phase spectra with increasing input power for
external flux biases of $\Phi_{dc} = 0$, $0.3\,\Phi_o$ and
$0.5\,\Phi_o$. In the case when $\Phi_{dc} = 0$, the magnitude and
phase spectra evolve from a symmetric shape to being asymmetric
with a lower resonance frequency as the power of the drive is
increased. The lower resonance frequency indicates that the
effective Josephson inductance over an oscillating period is
{higher}. For higher levels of the input power, the magnitude
spectrum exhibits a discontinuity near the resonance frequency,
where the system jumps from the lower branch to the higher branch.
The phase spectrum also exhibits a discontinuity similar to the
magnitude spectrum. For $\Phi_{dc} = 0.5\, \Phi_o$, the asymmetry
is opposite to that of $\Phi_{dc} = 0$; the resonance frequency
increases with higher power, indicating that the overall effective
inductance is decreasing with increasing power of the drive. An
intermediate behavior is captured at $\Phi_{dc}=0.3\,\Phi_o$. As
the input power increases, the nonlinear magnitude spectrum first
bends towards the lower frequency side, then gradually evolves
into a characteristic shape with two discontinuities near the
resonance frequency, once when the magnitude is increasing and
once when the magnitude is decreasing. Similarly, the phase
spectrum also shows two discontinuities at the same frequency
locations, with a partial phase drop at each discontinuity.

\begin{figure}[h]
   \begin{center}
   \includegraphics[width=3.3in]{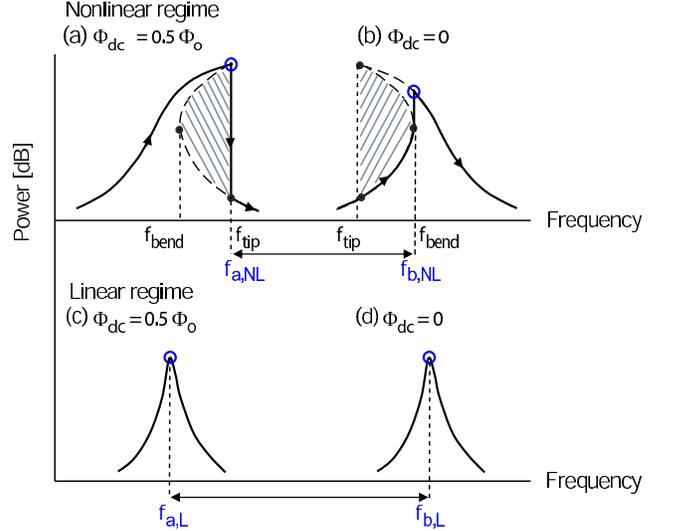}\\
   \caption{Illustration of the resonance spectra in the
   nonlinear regime (a and b) and the linear regime (c and d) for
   $\Phi = 0.5\,\Phi_o$ and $\Phi = 0$ respectively.
   The shaded region of the nonlinear spectrum marks the region over which multiple solutions occur.
The solid line traces the actual spectrum observed experimentally
with a forward frequency sweep, and the circular marker
corresponds to the peak frequency that was being measured. The
bending of the nonlinear spectra to opposite sides accounts for
the reduced separation of resonance frequencies
($f_{b,NL}-f_{a,NL}$) compared to the linear case
($f_{b,L}-f_{a,L}$), as was observed in
Fig.~\ref{fig:Fig2_modulation}b.}
   \label{fig:Bifur}
 \end{center}
\end{figure}

The shapes of these curves are similar to the response of driven,
weakly nonlinear systems which exhibit an instability region
indicating multiple solutions and hysteresis \cite{Strogatz,
Nayfeh}. Two such curves are shown in Fig.~\ref{fig:Bifur}a and b.
In particular, we model our system as a nonlinear circuit which
results from a current-driven LRC resonant circuit with a
nonlinear inductor $L$. In this case the flux in the inductor
$\Phi$ satisfies
\begin{equation}
I\sin\omega_s t =
C\frac{d^2\Phi}{dt^2}+\frac{1}{R}\frac{d\Phi}{dt}+ h(\Phi , d\Phi
/ dt)   \,, \label{eqndriving}
\end{equation}
where the function $h(\Phi , d\Phi / dt) $ models the nonlinearity
of the inductor.  For example, when $h=\Phi/L_o$ then the system
is a simple LRC resonant circuit with a linear inductor $L_o$.
When $h \sim \sin \Phi$ the nonlinear equation is analogous to a
driven pendulum system whose response is similar to
Fig.~\ref{fig:Bifur}b \cite{Strogatz, Barone}.  Another example is
the Duffing Equation where $h \sim \Phi- c\Phi^3$, whose response
is like Fig.~\ref{fig:Bifur}a for negative $c$ and like
Fig.~\ref{fig:Bifur}b for positive $c$ \cite{Strogatz, Nayfeh}. In
section~\ref{simsection} we will use a functional form for the
effective inductance which incorporates both the needed dependence
on applied DC flux and the resonance-frequency dependence observed
for small drives. In fact, given that
Fig.~\ref{fig:Fig2_modulation} shows that the resonance frequency
is periodic in the applied DC flux, then the effective inductance
that needs to be captured in the form of $h$ must also follow this
same periodicity.

We now use the illustration in Fig.~\ref{fig:Bifur} to explain
some of the general features of the data in
Fig.~\ref{fig:Fig2_modulation} and to motivate the subsequent
analysis. In Fig.~\ref{fig:Bifur}, the shape of the resonance
spectra are shown for both the linear and nonlinear regimes for
$\Phi_{dc} = 0$ and $0.5\,\Phi_o$, with the resonance frequency at
$\Phi_{dc} = 0.5\,\Phi_o$ lower than at $\Phi_{dc} = 0$ given the
flux-dependence of the effective inductance.  The shaded region of
the nonlinear spectrum corresponds to the region $f\in
[f_{tip},f_{bend}]$ over which multiple solutions occur (two of
which are stable and one of which is unstable)
\cite{NonlinearBook}. The solid line traces the actual spectrum
observed experimentally with a forward frequency sweep, and the
circular marker corresponds to the peak frequency that was being
measured.

We have seen from Fig.~\ref{fig:Fig2_modulation}b that as the
level of input power increases, the amount by which the frequency
is modulated over a flux quantum decreases.  This is a direct
consequence of the shape of the resonance spectrum as the system
response becomes increasingly nonlinear.  As illustrated in
Figs.~\ref{fig:Bifur}c and d, the resonance spectra at $\Phi = 0$
and $0.5\,\Phi_o$ have resonance frequencies that are maximally
separated ($f_{b,L}-f_{a,L}$) when the input bias is low, and
therefore when the resonance spectra are nearly those of a linear
response.  As the input current bias increases, the resonance
spectrum evolves from the symmetric Lorentzian shape to an
asymmetric shape. This is shown in Figs.~\ref{fig:Bifur}a and b.
The fact that the nonlinear spectra bend to opposite sides at
$\Phi=0$ and $0.5\,\Phi_o$ accounts for a reduced amount of
modulation in resonance frequency ($f_{b,NL}-f_{a,NL}$) compared
to the linear case.  It was also observed in
Fig.~\ref{fig:Fig2_modulation}b that the frequency modulation
curves for different input power meet periodically at the
inflection points, where the second derivative $d^2f_o/d\Phi^2$
equals zero.  In fact, the asymmetry of the spectrum changes sign
near the inflection points.

To further quantify the amount of bending in the nonlinear
spectrum, we introduce a parameter  $\delta f$ which is a
normalized shift of the resonance frequency $f_n$ of the nonlinear
spectrum relative to the linear spectrum $f_o$:
\begin{equation}
\label{eq:dw} \delta f= \frac{f_n-f_o}{f_o} \,.
\end{equation}
Experimentally, $f_o$ was determined as the resonance frequency of
the spectrum measured at the lowest power (-74\,dBm). $f_n$ was
defined as the peak frequency and, in the limit of high input
power, the frequency at which the discontinuity occurs.  The sign
of $\delta f$ serves as an indication of the polarity of the
bending. A {positive} $\delta f$ corresponds to the nonlinear
spectrum bending to the {higher} frequency side, and a {negative}
$\delta f$ corresponds to the spectrum bending to the {lower}
frequency side.

\begin{figure}[h]
   \begin{center}
   \includegraphics[width=3.5in]{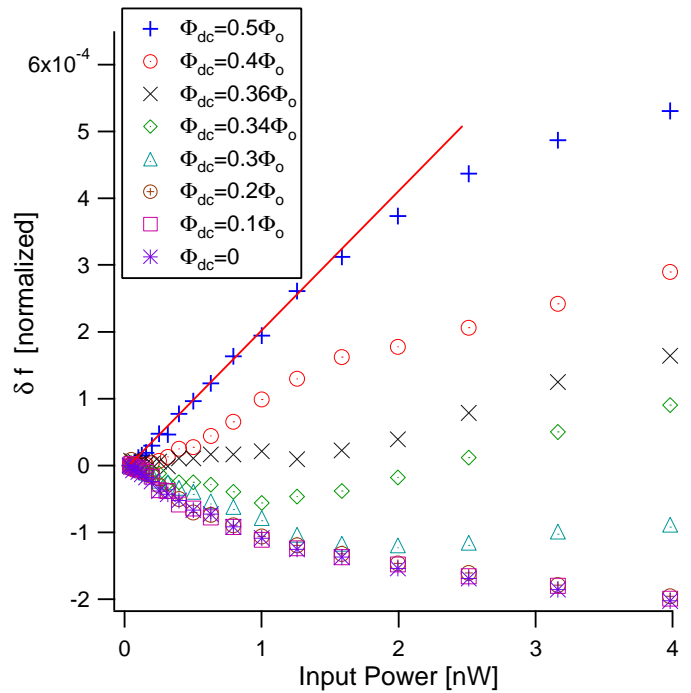}\\
   \caption{(Color online) Bending $\delta f$ as a function of input power from -74\,dBm to -54\,dBm in 1\,dB steps.
   Measurements were made with a forward frequency sweep.
   The data are shown for various DC flux biases between 0 to $0.5\,\Phi_o$.
   The sign of $\delta f$ indicates the polarity of the
   asymmetric spectrum.  The red line corresponds to a linear
fit in the low-power regime.}
   \label{fig:Fig4_AmtBend_July1}
 \end{center}
\end{figure}

In Fig.~\ref{fig:Fig4_AmtBend_July1} the normalized frequency
$\delta f$ of the resonance spectrum is plotted for increasing
input power from -74\,dBm to -54\,dBm.  The measurements of the
spectra were made with a forward frequency sweep.  The different
markers correspond to various flux biases between $\Phi_{dc} = 0$
to $0.5\,\Phi_o$. At $\Phi_{dc}=0.5\,\Phi_o$ (top plot), $\delta
f$ is increasingly {positive}; whereas,  at $\Phi_{dc} = 0$
(bottom plot), $\delta f$ becomes increasingly {negative}.
Furthermore, the amount of bending $|\delta f|$ at $\Phi_{dc} = 0$
is smaller than at $0.5\,\Phi_o$ for a given input bias, which is
related to the fact that a forward frequency sweep captures the
full frequency extent of the bistable region for
$\Phi_{dc}=0.5\,\Phi_o$ but not for $\Phi_{dc}=0$. (The reverse is
true if the frequency is swept backwards, as discussed in the next
section.) Finally at intermediate flux biases between
$0.3\,\Phi_o$ to $0.36\,\Phi_o$, $\delta f$ shows an undulating
behavior, corresponding to the asymmetric spectrum constantly
varying its polarity. There is an initial linear dependence of
$\delta f$ on input power, which will be discussed in the next
section.

\section{Hysteresis of the Resonance Spectrum} \label{sect:Hystersis}

For larger input drives, the resonance spectrum exhibits a
discontinuity which corresponds to one of the two boundaries of
the bistable region. Within the bistable region, the system
settles into one of the solutions depending upon the initial
conditions. For our case, the initial condition is set by the
solution at the previous driving frequency, which in turn is
determined by the direction of the frequency sweep.  The resonant
behavior of the readout circuit presented so far were obtained
with a forward frequency sweep. Here, we present the hysteretic
behavior of the resonance spectrum measured with both forward and
backward frequency sweeps so that the full boundary of the
bistable region can be mapped.

\begin{figure}[ht]
   \begin{center}
   \includegraphics[width=3.5in]{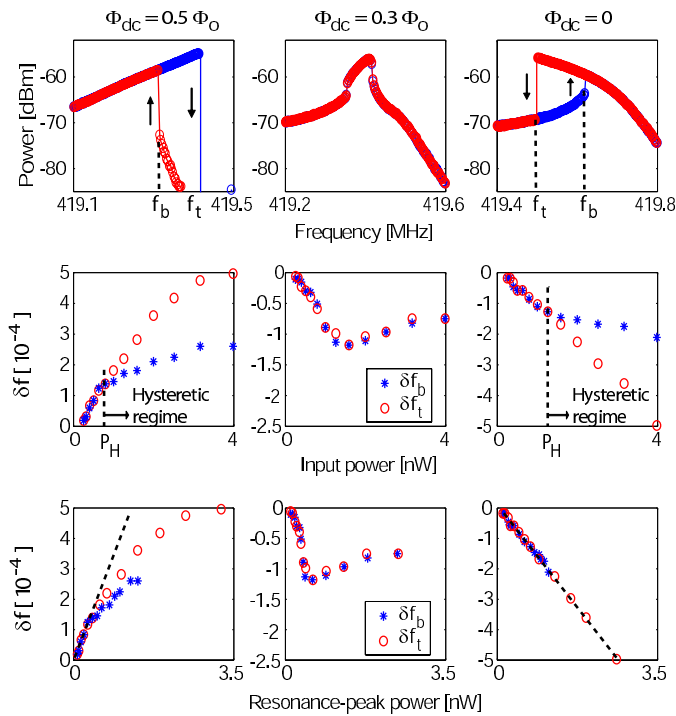}\\
   \caption{(Color online) Top plots: Hysteretic resonance spectrum for flux biases at $\Phi_{dc}=0$, $0.3\,\Phi_o$
   and $0.5\,\Phi_o$.  The extent of the bistable region is given by
   $|f_t-f_b|$.  Middle plots: $\delta f_t$ and $\delta f_b$ as a function
   of input power from -66\,dBm to -54\,dBm in 1\,dB steps.  Onset of
   hysteretic regime occurs at $P_H$=-61\,dBm for
   $\Phi_{dc}=0.5\,\Phi_o$ and $P_H$=-59\,dBm for $\Phi_{dc} = 0$.
   Bottom plots: $\delta f_t$ and $\delta f_b$ re-plotted as
   a function of resonance-peak power.  The dotted line is a linear fit for low power.}
   \label{fig:Fig5_May24_Hys_SnapShots}
   \end{center}
\end{figure}

The top three plots in Fig.~\ref{fig:Fig5_May24_Hys_SnapShots}
show the typical hysteretic spectrum for flux biases at
$\Phi_{dc}=0$, $0.3\,\Phi_o$ and $0.5\,\Phi_o$.  The data are
shown for an input power level of -54\,dBm, which corresponds to a
highly nonlinear regime. The direction of the frequency sweep is
indicated by the arrows.  For the case of $\Phi_{dc}=0$ and
$0.5\,\Phi_o$, we define the extent of the bistable region as
$|f_t-f_b|$, where $f_b$ is the frequency at which the resonance
spectrum jumps from the lower to the higher stable branch, and
$f_t$ corresponds to the frequency at which the spectrum falls
from the higher to the lower stable branch. At $\Phi_{dc} =
0.3\,\Phi_o$, the forward and backward traces overlapped,
indicating that the bistable region associated with the two
discontinuous edges were too small to be detected given the
frequency resolution.

The onset of the hysteretic regime is illustrated in the middle
three plots of Fig.~\ref{fig:Fig5_May24_Hys_SnapShots}, where the
extent of the bistable region was characterized as a function of
input power from -66\,dBm to -54\,dBm in 1\,dB steps.  First, we
normalized $f_t$ and $f_b$ with respect to the resonance frequency
$f_o$ in the linear regime according to a definition similar to
Eqn.~\ref{eq:dw}:
\begin{equation}
\delta f_t = \frac{f_t-f_o}{f_o} \qquad \mbox{and} \qquad \delta
f_b = \frac{f_b-f_o}{f_o}
\end{equation}
$\delta f_t$ and $\delta f _b$ were then plotted as a function of
input power.  Hysteretic behavior was observed when the input bias
was above a threshold $P_H$, which was measured to be -61\,dBm for
$\Phi=0.5\,\Phi_o$, and at a higher power of -59\,dBm for
$\Phi=0$.

In the bottom row of plots of
Fig.~\ref{fig:Fig5_May24_Hys_SnapShots}, we have plotted $\delta
f_t$ and $\delta f_b$ as a function of the resonance-peak power.
We see that $\delta f_t$ is initially a linear function of the
resonance-peak power for all three fluxes shown and that it is
fully linear for zero DC flux. This initial linear dependence on
resonance peak power is found for many functional forms of the
nonlinear term $h(\Phi, d\Phi / dt)$ \cite{Nayfeh, Strogatz}.
However, the dependence on input power varies according to the
particular functional form of $h(\Phi, d\Phi / dt)$, see reference
\cite{Nayfeh} for some sample cases. For example, both the Duffing
equation and the pendulum model give a linear dependence on both
input and resonance-peak powers in the weakly nonlinear regime
\cite{Landau_book,Nayfeh,NonlinearBook}. For $ h = \Phi (d\Phi /
dt)^2$, $\delta f_t$ depends linearly on the resonance-peak
amplitude, but as the cube root of the input power \cite{Nayfeh}.

\section{Simulations of nonlinear resonant behavior based on a
phenomenological model} \label{simsection}

In this section, we present simulations to illustrate the
nonlinear resonant behavior of the readout circuit.  One approach
to simulate the AC-driven behavior of a circuit comprising a SQUID
is to numerically solve the set of coupled differential equations
governing the SQUID consistently with the rest of the circuit.
However, the dynamical modeling of the resulting circuit is
complex; for example, for a circuit with a SQUID shunted by a
resonating capacitor has 6 dynamical variables when the mutual
inductive coupling between the SQUID and the resonating loop is
included \cite{Jan_PhDthesis}.
Therefore, we use the phenomenological LRC circuit model of
Eqn.~\ref{eqndriving}  with the linear inductance replaced by a
flux-dependent nonlinear inductor. This approach allows a
reduction of the mathematical complexity of the problem to one
dynamical variable.

\begin{figure}[ht]
   \begin{center}
   \includegraphics[width=3.5in]{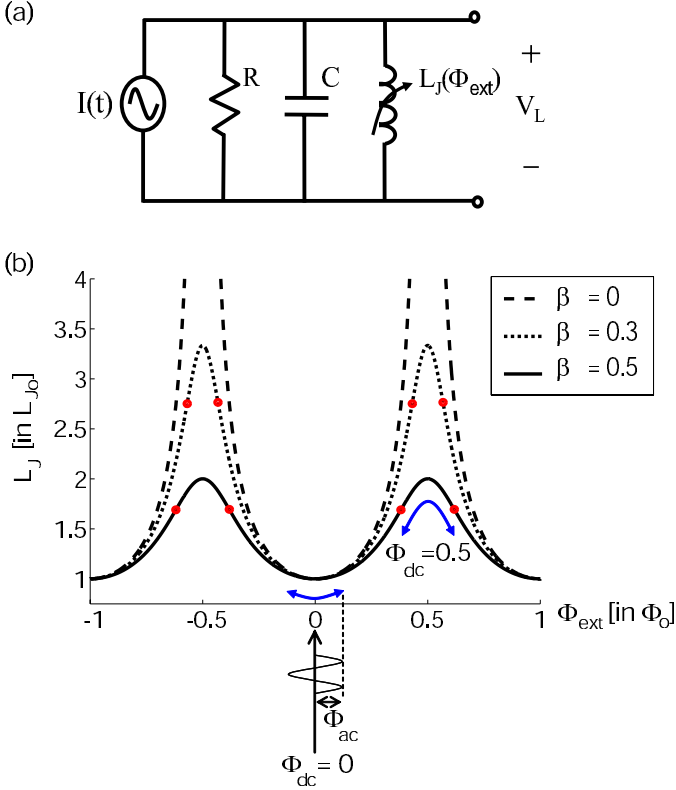}\\
   \caption{(a) Circuit schematic of the phenomenological resonant model. (b)
   A plot of $L_J(\Phi_{ext})$ given by Eqn.~\ref{eq:LJ_toy}
   for different values of $\beta$.  The circular markers represent the inflection points where
   ${d^2L_J}/{d\Phi^2}$ is zero.  The illustration shows that depending on the DC flux bias,
   the AC modulation of $L_J$ due to $\Phi_{ac}$
   can result in a lower ($\Phi_{dc}=0.5\,\Phi_o$) or higher ($\Phi_{dc} = 0$) effective inductance
   .}
   \label{fig:Fig6_Toymodel}
   \end{center}
\end{figure}

Specifically, the readout SQUID is modeled by a flux-dependent
nonlinear inductor $L_J(\Phi_{ext})$ given by
\begin{equation}
L_J(\Phi_{ext}) = \frac{L_o}{ \sqrt{(1+ \beta^2) + (1- \beta^2)
\cos({2\pi\Phi_{ext}}/{\Phi_o}) }} \,. \label{eq:LJ_toy}
\end{equation}
The functional form for the nonlinear inductor $L_J(\Phi_{ext})$
in Eqn.~\ref{eq:LJ_toy} captures the sinusoidal-like shape of the
frequency response of the actual readout circuit as previously
shown in Fig.~\ref{fig:Fig2_modulation}b.  This is also
illustrated in Fig.~\ref{fig:Fig6_Toymodel}b, where
$L_J(\Phi_{ext})$ is plotted for $L_o = \sqrt{2}{L_{Jo}}$
(Eqn.~\ref{eq:LJo}), and for different values of $\beta$. It can
be seen that $\beta$ has an effect on (a) the amount by which
$L_J$ is modulated over half a flux quantum, and (b) the position
of the inflection points at which the second derivative
${d^2L_J}/{d\Phi^2}$ is zero. One can therefore use $\beta$ in
Eqn.~\ref{eq:LJ_toy} as a fitting parameter such that the location
of the inflection points of the frequency response
(${d^2f_o}/{d\Phi^2} =0 $) match the data in
Fig.~\ref{fig:Fig2_modulation}b. The functional form of the
inductance was motivated by that of an asymmetric SQUID
\cite{Barone} since it has the needed periodicity; however, it is
only the form that is used, and we are not assuming that asymmetry
plays any role in the physical circuit.

We further assume that the inductor is mutually coupled to a total
external flux bias as $\Phi_{ext} = \Phi_{dc}+\Phi_{ac}$, where
$\Phi_{dc}$ is the DC flux bias that was applied experimentally to
the SQUID, and $\Phi_{ac}$ is any oscillating flux that was
mutually coupled to the SQUID. This AC-part of the flux is modeled
as being proportional to the self-induced flux as $ \Phi_{ac}
=\alpha\Phi$ to ensure that the amount of coupled flux increases
as $\Phi$ gets large near the resonance frequency of the circuit.
The size of $\Phi_{ac}$ used in the simulations is around
$0.1\,\Phi_o$ near the resonance frequency.

The dynamics of the phenomenological resonant circuit is then
governed by Eqn.~\ref{eqndriving} and
\begin{equation}
h  = \frac{\Phi}{L_J(\Phi_{ext})} \label{hsimulations}
\end{equation}
For small drives ($\alpha \approx 0$), this term can be
approximated by a linear inductance which depends on $\Phi_{dc}$
in a sinusoidal-like fashion. For larger values of the drive,
Eqn.~\ref{hsimulations} can be expanded to be in the form of
$a\,\Phi+ b \, \Phi^2 + c \, \Phi^3$ when $
(1-\beta^2)/(1+\beta^2)<< 1$.  Moreover, $b \sim \sin{(2\pi
\Phi_{dc}/\Phi_o)}$ and $ c \sim \cos{(2\pi \Phi_{dc}/\Phi_o)}$.
For $\Phi_{dc} = 0$ and $\Phi_{dc} = 0.5\,\Phi_o $,  then $b = 0$,
and the resulting equation is of the form of a Duffing Equation
with both a linear and cubic term in $\Phi$, so that with
increasing drive there will be a bending of the resonance
frequency with its associated hysteresis \cite{Nayfeh} . Moreover,
the sign of the cubic term is opposite for $\Phi_{dc} = 0$ and
$\Phi_{dc} = 0.5\, \Phi_o$, and hence bending will be in opposite
directions, as needed to qualitatively explain the data. At other
values of $\Phi_{dc}$ there will be a competition between the
quadratic and cubic terms.

\begin{figure}[ht]
   \begin{center}
   \includegraphics[width=3.5in]{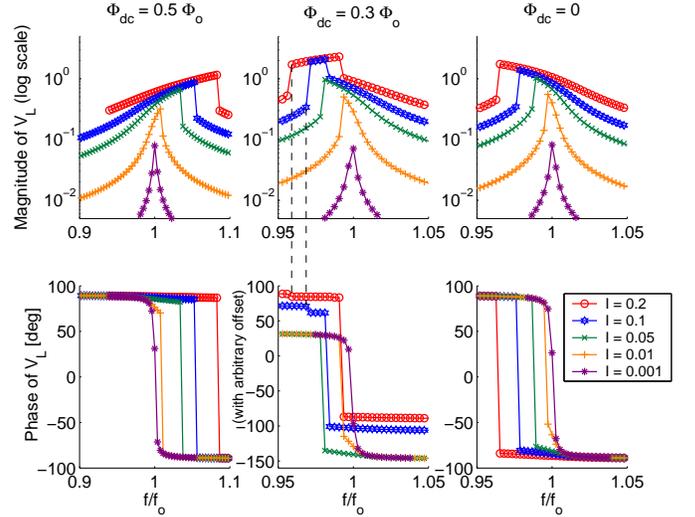}\\
   \caption{(Color online) Simulated magnitude and phase spectra of $V_L$
   for increasing drive amplitude $I$.  The
   results
   qualitatively reproduce the experimentally observed behavior in
   Fig.~\ref{fig:Fig3_Mag_Phase_compareToy}.  $V_L$ is plotted in
   reduced units of $\Phi_o/\sqrt{L_{Jo}{C}}$ and $I$ in units of $2I_{co}$.  The
   frequency axes are normalized with respect to the resonance frequency of the linear
   spectrum:
   $f_o$ = 1.88\,GHz ($\Phi_{dc}=0$), 1.59\,GHz ($0.3\,\Phi_o$) and 1.33\,GHz ($0.5\,\Phi_o$).
   The phase spectra at the highest drives I = 0.1 and 0.2 for
   $\Phi_{dc} = 0.3\,\Phi_o$ are arbitrarily shifted for display
   purpose.
   }
   \label{fig:Fig7_ToyModelSim}
   \end{center}
\end{figure}

By numerically solving Eqn.~\ref{eqndriving} for $\Phi$ at
different driving frequencies $\omega_s$, the magnitude and phase
spectra of the voltage across the inductor $V_L = d\Phi/dt$ were
obtained.  In Fig.~\ref{fig:Fig7_ToyModelSim} the spectra are
shown for increasing drive amplitude for $\Phi_{dc}=0$,
$0.3\,\Phi_o$ and $0.5\,\Phi_o$. $V_L$ is plotted in reduced units
of $\Phi_o/\sqrt{L_{Jo}{C}}$, and the drive amplitude $I$ is in
units of $2I_{co}$. The driving frequency was swept such that the
lower stable branch within the bistable region is shown for all
flux biases. The simulation qualitatively resembles the
experimental data presented in
Fig.~\ref{fig:Fig3_Mag_Phase_compareToy}.  As expected, the
nonlinear spectrum has a lower resonance frequency at
$\Phi_{dc}=0$, and a higher resonance frequency at
$\Phi_{dc}=0.5\,\Phi_o$. Also, discontinuities are observed at the
boundary of the bistable region for the higher biases. At
$\Phi_{dc}=0.3\,\Phi_o$, the magnitude spectra at the two highest
input biases exhibit two discrete jumps, once at a lower frequency
when the magnitude is increasing, and another at a higher
frequency when the magnitude is decreasing.  As for the phase
spectra, a partial phase drop occurs at the low-frequency
discontinuity, while most of the phase drop occurs at the
high-frequency discontinuity.

We have also performed simulations which qualitatively reproduce
the hysteresis data in Fig.~\ref{fig:Fig5_May24_Hys_SnapShots}.
This was done by stepping the driving frequency in both the
low-to-high and high-to-low frequency directions, and by ensuring
the initial conditions used for the next frequency point were the
solutions obtained for the previous frequency point. We have
assumed that the square of the drive amplitude $I^2$ for the
simulations is proportional to the input power for the experiment.

\begin{figure}[tbh]
   \begin{center}
   \includegraphics[width=3.5in]{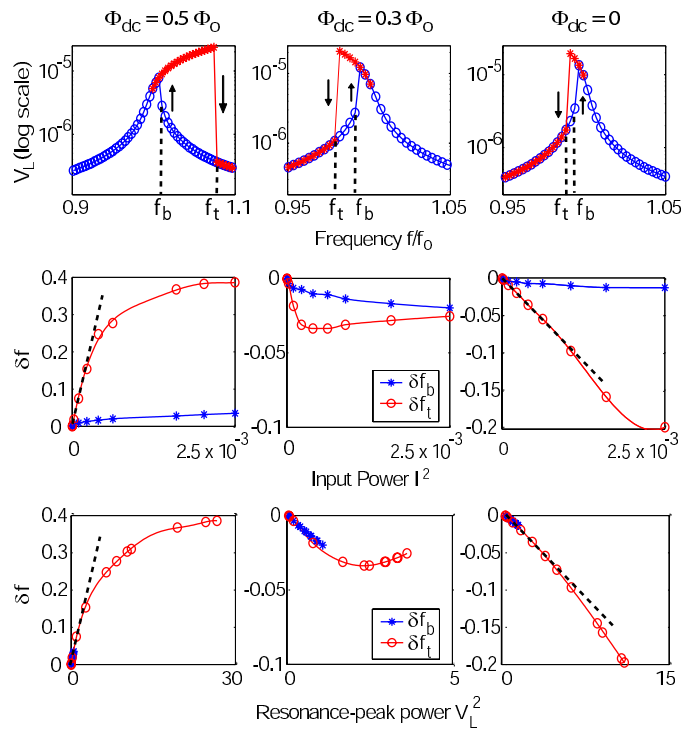}\\
   \caption{(Color online) Top row: simulated hysteretic behavior at various flux bias
   for drive amplitude $I= 0.01$. Middle row: $\delta f_b$ and
   $\delta f_t$ as a function of the square of the drive amplitude, which is proportional to the input power.
   The simulations were performed for $I$ between 0.001 to 0.05.
   Bottom row: $\delta f_b$ and
   $\delta f_t$ as a function of the square of the voltage response $V_L$,
   which is proportional to the resonance-peak power.  $V_L$ is in
   reduced units of $\Phi_o/\sqrt{L_{Jo}{C}}$ and $I$ is in units of $2I_{co}$.}
   \label{fig:Fig8_ToyHys}
   \end{center}
\end{figure}

The top row of plots in Fig.~\ref{fig:Fig8_ToyHys} shows the
typical simulated hysteretic behavior at various flux biases for
$I/2I_{co}=0.01$.  The extent of the bistable region given by
$|\delta f_b$-$\delta f_t|$ is the largest at $\Phi_{dc} =
0.5\,\Phi_o$ for this drive amplitude. The middle and bottom rows
of plots show the dependence of $\delta f_b$ and $\delta f_t$ as a
function of input power and resonance-peak power respectively. The
trend at $\Phi_{dc} = 0.3\,\Phi_o$ shows qualitative resemblance
to the experimental data.  The magnitude of $\delta f_t$ initially
increases linearly with the power of the drive for $\Phi_{dc}=0$
and $0.5\,\Phi_o$ similar to the data. As explained previously,
this linear dependence is expected at low drives due to the nature
of the nonlinearity \cite{Landau_book, Nayfeh}.

The phenomenological model presented here is meant to show the
qualitative trends in the data. To be more quantitative, we have
analyzed more complex circuits \cite{Jan_PhDthesis}. For example,
we have considered the current-driven circuit across an asymmetric
SQUID with self-inductance, and mutual inductive coupling between
the SQUID loop and the resonating loop. The induced flux in the
resonating loop is enhanced by the quality factor Q near the
resonance frequency. The simulation gives results which
qualitatively reproduce the data with reasonable numbers; however,
the quantitative fitting of the data was not possible due to the
uncertainty in the actual on-chip values for the capacitances and
the mutual/self inductances.

\section{Discussion}

In this paper, we characterized the resonant behavior of the
readout circuit to be utilized in a resonant scheme for measuring
a PC qubit. We have identified the linear, nonlinear (bistable),
and  hysteretic (highly nonlinear) bias regimes of operation.
Given the high quality factor of the resonance, we observed a
manifestation of the nonlinearity due to the Josephson inductance
of the readout SQUID. The resonance spectrum of the readout
circuit became asymmetric in the nonlinear regime, and the
polarity of the asymmetry changed sign as a function of DC
magnetic flux bias to the SQUID.  The numerical simulations using
a nonlinear inductor qualitatively reproduce the trends in the
experimental data.

To perform time-resolved measurements of the qubit on a
microsecond timescale, the resonant readout is to be operated at a
bias frequency $f_s$ near the resonance frequency $f_o$.  The
change in the resonance frequency due to the qubit signal is thus
detected as a difference in the magnitude or phase of the output
voltage at $f_s$.

Finally, it should be mentioned that while operating the resonant
readout in the linear regime keeps the input bias low and reduces
the level of decoherence on the qubit, the readout operated in the
nonlinear regime has the advantage of being used as a bifurcation
amplifier \cite{Siddiqi}. In particular, the bias frequency $f_s$
can be chosen within the bistable region of the nonlinear spectrum
such that the system has two stable solutions corresponding to
different voltages. The probability of occupancy in the
higher/lower stable solution is  sensitive to changes in the
resonance frequency $f_o$ (qubit-mediated) relative to $f_s$, and
thus provides additional sensitivity for qubit readout over the
linear approach.

\section{Acknowledgements}

We thank Daniel Oates, Yang Yu, Jonathan Habif, Sergio Valenzuela
for useful discussions, and Terry Weir for technical assistance.
This work was supported in part by AFOSR Grant No.
F49620-01-1-0457 under the DURINT Program, and by an NSF graduate
fellowship. The work at Lincoln Laboratory was supported by the US
DOD under Air Force Contract No. F19628-00-C-0002.

\end{document}